# Saturated absorption competition microscopy

Guangyuan Zhao[1], Mohammad M Kabir[2], Kimani C. Toussaint Jr.[3], Cuifang Kuang[1*],
Cheng Zheng[1], Zhongzhi Yu[1] & Xu Liu[1*]

[1]State Key Laboratory of Modern Optical Instrumentation, College of Optical Science and Engineering, Zhejiang University, Hangzhou 310027, China

[2]Department of Electrical and Computer Engineering, University of Illinois Urbana-Champaign, Urbana, Illinois 61801, USA.

[3]Department of Mechanical Science and Engineering, University of Illinois Urbana-Champaign, Urbana, Illinois 61801, USA.

*Correspondence and requests for materials should be addressed to C.K. (email: cfkuang@zju.edu.cn) and X.L. (email: liuxu@zju.edu.cn).

**ABSTRACT:** We introduce the concept of saturated absorption competition (SAC) microscopy as a means of providing sub-diffraction spatial resolution in fluorescence imaging. Unlike the *post*-competition process between stimulated and spontaneous emission that is used in stimulated emission depletion (STED) microscopy, SAC microscopy breaks the diffraction limit by emphasizing a *pre*-competition process that occurs in the fluorescence absorption stage in a manner that shares similarities with ground-state depletion (GSD) microscopy. Moreover, unlike both STED and GSD microscopy, SAC microscopy offers a reduction in complexity and cost by utilizing only a single continuous-wave laser diode and an illumination intensity that is ~ 20x smaller than that used in STED. Our approach can be physically implemented in a confocal microscope by dividing the input laser source into a time-modulated primary excitation beam and a doughnut-shaped saturation beam, and subsequently employing a homodyne detection scheme to select the modulated fluorescence signal. Herein, we provide both a physico-chemical model of SAC and experimentally demonstrate by way of a proof-of-concept experiment a transverse spatial resolution of ~$\lambda/6$.

Owing to the advantage of being non-invasive in observing living samples, far-field optical microscopy is widely used in the life sciences. However, the existence of the diffraction barrier leads to the poor imaging of samples with spatial features smaller than approximately half the wavelength of the probes. Indeed, the discovery by Abbe in 1873 reveals that the smallest focal spot achievable by a lens is restricted by diffraction to approximately $\lambda/2NA$ [1], where $\lambda$ denotes the wavelength of light and NA is the numerical aperture of the focusing lens. It was not until the 1990s that the first viable principle to break the diffraction barrier appeared[2, 3], thanks in no small part to the progress in fluorescence probe technology. Since that time, various techniques have been proposed to enhance the spatial resolution in fluorescence microscopy [2, 4-8].

In the development of super-resolution fluorescence microscopes, the saturation phenomenon has been widely used to induce a strong nonlinearity in the excitation and fluorescence emission rates. The resulting smaller effective point spread function (with correspondingly higher spatial-frequency components) leads to the enhancement in spatial resolution. Saturated excitation (SAX) microscopy applies the nonlinear response in the saturation of the fluorescence molecule excited stage [7] to extract high spatial-frequency components, thereby obtaining structural information beyond the diffraction-limited resolution. Stimulated emission depletion (STED) [2], along with other reversible saturable (or switchable) optical linear (fluorescence) transitions (RESOLFT) [9] microscopy techniques, exploits stimulated emission to suppress the spontaneous fluorescence emission and further depresses it by the saturation effect [10]. The resolution of STED has been demonstrated down to 2.4 nm [11]. Because STED can obtain nano-scale superresolution, this approach has been more widely used and explored.

In the STED process, a doughnut beam switches the molecules off by forcing them down to the $S_0$ electronic ground state [12], whereby the resulting competition between stimulated emission and spontaneous decay yields a fluorescence signal with a sub-diffraction point spread function. Herein we refer to this procedure as *post*-competition because it occurs in the fluorescence emission stage, whereas *pre*-competition occurs in the fluorescence absorption (excitation) stage. This analogy is also observed in ground state depletion (GSD) [13, 14], a similar but not identical photo-switching based microscopy technique. In GSD the fluorophores are "switched off" by transiently shelving the fluorophore in its metastable dark triplet state $T_1$. In addition, two other super-resolution approaches have used absorption saturation. Wang et al. introduced a third beam into pump-probe methods enabling super-resolution imaging of non-fluorescent chromophores via the 'saturated transient absorption' [15], with the constraint of more complex and expensive hardware requirements. Yang et al. introduced excitation stage saturation and super-resolved single molecules by demodulating the overlapping optical resonance frequencies [16]. However, their approach required cooling at liquid helium temperatures (2K) and specialized single fluorescent molecules with narrow zero-phonon lines. Thus, demonstration of sub-diffraction fluorescence imaging using a single low-powered continuous-wave laser source that exploits competition at the fluorescence absorption stage and at room temperature would be accessible and appealing to a broad audience of researchers.

In this letter, we present a *pre*-competition method that utilizes the competition between two incident beams derived from the same continuous-wave (CW) source at the absorption stage, to develop saturated absorption competition (SAC) microscopy. SAC offers reduced complexity and cost, and is developed by inducing competition during the fluorescence absorption process and separating the emitted mixed fluorescence signals by homodyne detection. We provide a theoretical physico-chemical model of SAC and experimentally verify its effectiveness by imaging sub-wavelength fluorescent particles with a transverse spatial resolution of ~100 nm (~$\lambda$/6). To the best of our knowledge, our approach is the first demonstration of superresolution imaging resulting from an engineered competition in the fluorescence absorption stage at room temperature using a single CW laser source with an illumination intensity that is ~20x smaller than that used in conventional STED microscopy.

*Physico-chemical model of SAC.* To understand the principle of SAC, we first consider the relationship between the excitation and the fluorescence intensities with regard to the effect of saturation. The essence of the excitation-emission process is based on a photo-physical model, which can be observed from the energy diagram depicted in Fig. 1(a) [17], where $S_0$ represents the ground singlet state, $S_1$ the first-excited singlet states, $T_1$ the lowest-excited triplet state, and $S_n$ and $T_n$ represent the higher excited singlet and triplet states, respectively (more details are provided in **Supplemental Material, Note 1**). The emitted fluorescence signal is proportional to the population of fluorescence molecules in $S_1$. We consider Rhodamine 6G as the fluorescent molecule in our model. The dependence of the emitted fluorescent intensity on the input excitation intensity is determined from the equations given in **Supplemental Note 1.** Fig. 1(b) shows the relationship between the excitation and emission intensities. Under non-saturation situations, the emission intensity is proportional to the excitation intensity, as shown by the black line in Fig. 1(b). Conversely, under saturation the relationship between these intensities becomes nonlinear [blue curve in Fig. 1(b)]. This nonlinearity degrades the image equality, and as a result absorption (excitation) saturation by itself has been traditionally avoided in laser scanning microscopy. However, it is now well-understood that the use of a doughnut-shaped intensity saturation beam in conjunction with a lower-intensity Gaussian beam has the benefit of significantly reducing the spatial extent of the latter, thereby increasing the spatial resolution beyond the diffraction limit. To further demonstrate this concept, let us consider two beams of the same wavelength that are incident on a fluorescent sample as shown by the arrows marked A (with solid lines) and B (with dashed lines) in Fig. 1(c). When the sum intensity of these two beams is relatively too weak to produce saturation as shown in the left subfigure of Fig. 1(c), $S_1$ remains sufficiently unoccupied and the probability of emission of a photon is constant. As we continue increasing the intensity of A and pass a certain threshold, the accompanied saturation of molecules in the fluorescence absorption stage leads to a drop in the emission probability, which reduces the emission intensity of beam B, as shown in the right subfigure in Fig. 1(c). The rate constants for the first and second beams are given by $k_{exc}$ and $k_{exc'}$ respectively. Thus, combining with Eq. (S2-a) **Supplemental Note 1,** we derive the probability of $S_1$, under saturated conditions with dual excitation from two identical beams, as

$$S_{1eq} = \frac{k_T(k_{exc}+k_{exc'})}{(k_{exc}+k_{exc'})(k_{isc}+k_T)+k_0 k_T}.$$

(1)

Using the values for Rhodamine 6G for $k_T$ ($4.9*10^5$ s$^{-1}$), $k_0$ ($2.56*10^8$ s$^{-1}$), and $k_{isc}$ ($1.1*10^6$ s$^{-1}$) [17], Eq. (1) can be further simplified to

$$S_{1eq} = (k_{exc'}+k_{exc})/(3.245\times(k_{exc'}+k_{exc})+k_0). \tag{2}$$

Proportioned by the intensities of two beams, the effective probability of the first beam at $S_1$ will be expressed as

$$S_{1eff} = k_{exc} / (3.245(k_{exc'} + k_{exc}) + k_0), \qquad (3)$$

which is similar to that of STED, $S_{1eff} = k_{exc} / (3.245 \times k_{exc} + k_{STED} + k_0)$ [18]. As we have pointed out, sub-diffraction resolution for both STED and SAC processes are obtained by depressing the effective probability with a competition, the former's post-competition corresponds to adding an extra decay rate $k_{STED}$ to compete with spontaneous emission, while the latter's pre-competition corresponds to splitting the same beam and adding another excitation rate $k_{exc'}$ to introduce competition under fluorescence absorption. In practice, the addition of the second beam with high intensity renders $k_{exc'}, k_{STED} \gg k_{exc}$, and the effective probability in $S_1$ relies on the numerical relationship of $k_{exc'}$ and $k_0$. As is often done in STED, we take $I_s$ as the equivalent saturation intensity, given by $k_0 = \sigma_{01} I_s \lambda / hc_1$, the value when the effective probability of $S_1$ drops down to half of the initial value with competition.

Considering that $k_{exc}$ is multiplied by a factor of 3.245 and that $\sigma_{01}$ (2.7*10^-16 cm$^2$) is typically nine folds that of $\sigma_{sted}$ (3*10^-17 cm$^2$) [18], the rate of decrease in SAC is notably faster than STED, which can also be seen in the Fig.1(d). In this simulation model of Rhodamine 6G, $I_{SC}$ (167 kW/cm$^2$) is about 20 folds smaller than $I_{SD}$ (3.3 MW/cm$^2$), coinciding with the value of the conventional CW STED beam calculated by Willig et al [18], meaning that the required illumination intensity to render the same spatial resolution by SAC is much smaller than that of STED.

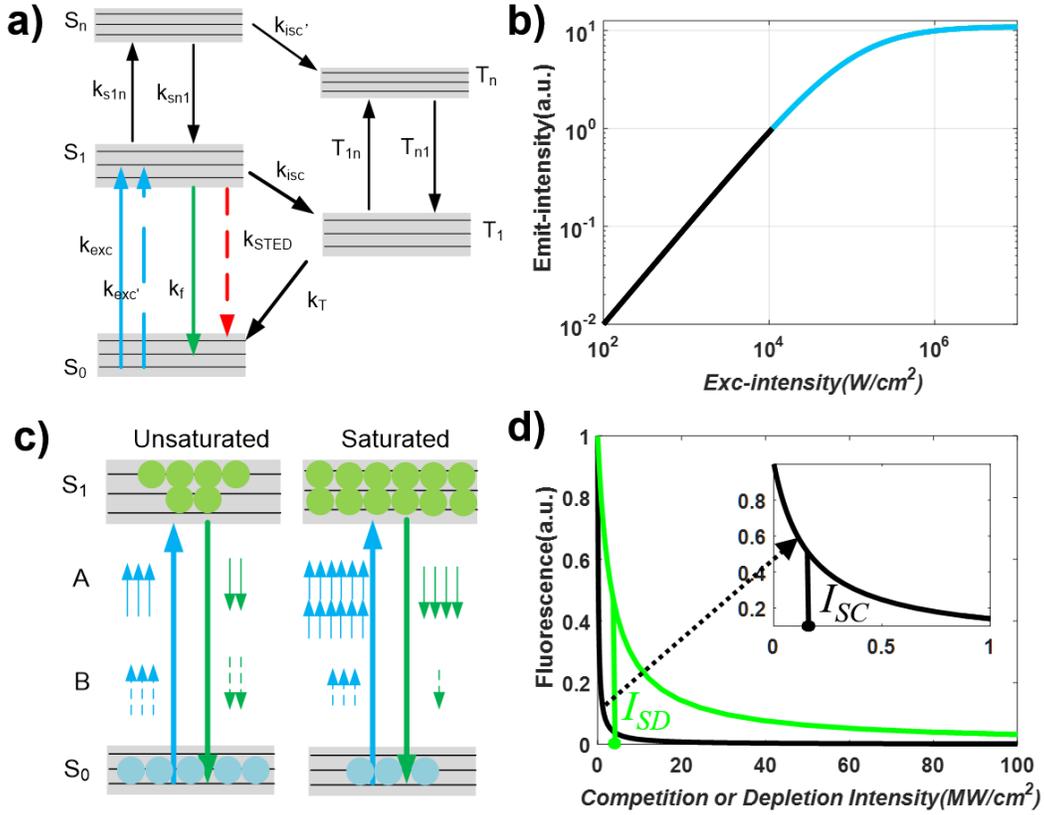

**FIG. 1. Principle model of SAC.** **(a)** Five-level molecular electronic state model used to calculate the relationship between the excitation-emission intensities. **(b)** Saturation effect in fluorescence emission from rhodamine 6G molecules calculated using the five-level electronic model. **(c)** Illustration of the saturation of $S_1$ state and the process of competition during the excitation. 'A' denotes the high-intensity competition beam (solid lines) while 'B' denotes the low-intensity primary excitation beam (dashed lines). **(d)** Decrease of the normalized fluorescence value ($I_s$) in the $S_1$ state of SAC (black curve) and STED (green curve) as a result of competition. To differentiate $I_s$ of the two methods, we take $I_{SC}$ as the $I_S$ of SAC and $I_{SD}$ as that of STED, which are represented by the green and black (inset) lines in (d), respectively.

*Imaging process of SAC microscopy.* Based on the principle outlined in the previous sub-section, SAC microscopy is designed to break the diffraction limit through absorption competition in a time-modulated confocal microscope by adding a doughnut saturation beam with the same wavelength as the primary beam but with much higher intensity. As presented in Fig. 2, a home-built confocal microscope is modified to accommodate two beams derived from a laser source. A polarized beam splitter (PBS) is used to split the incident beam into two orthogonally polarized beams. The first beam is phase modulated by a 0-2π vortex phase plate (VPP) to generate a doughnut beam [shown in Fig. 2(a)] with illumination intensity hundreds of times larger than that used in a traditional confocal system. As a result, absorption saturation of the fluorophore occurs and a large amount of fluorescence is emitted from the peripheral region of the fluorophore. The second beam is a standard Gaussian beam [shown in Fig. 2(b)] and is introduced at a relatively lower illumination intensity. An acousto-optic modulator (AOM) is used to time modulate this beam at a certain frequency to excite the geometric center of the fluorophore and subsequently permit sensitive coherent discrimination from the fluorescence generated from the unmodulated doughnut beam. Consequently, two overlapping beams excite the sample simultaneously, leading to a composite excited spot as shown in Fig. 2(c). When the intensity of doughnut illumination light increases and induces saturation, the outer fluorescence emission PSF becomes flat and wide, narrowing the center region of the doughnut emission spot. Subsequently, much more fluorescence is emitted by the doughnut mode of light and only the solid spot near the very center of the focus emits fluorescence, narrowing the fluorescence emission PSF of the solid mode. Finally, a sub-diffraction PSF [shown in Fig. 2(d)] is obtained by detecting the acquired fluorescence emission with a lock-in amplifier, which synchronizes the frequency to the AOM through a signal generator (see details in **Supplemental Material Note 2, Supplemental Material Fig. 1 and Supplemental Material Fig. 2**).

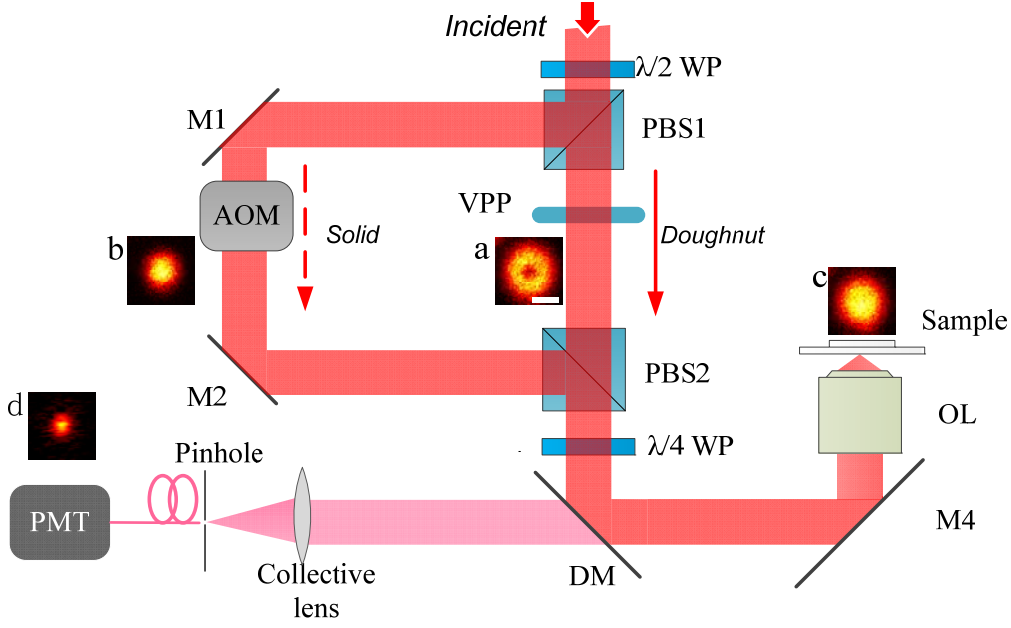

**FIG. 2. Conceptual schematic of saturated absorption competition microscopy.** PMT, photomultiplier; DM, dichroic mirror; OL, objective lens; M, reflective lens; WP, wave plate. Solid arrow and dashed arrow represent the sub-beams with phase modulation and the other with frequency modulation, respectively. The $\lambda/2$ WP is used here to tune the intensity ratio of the two beams. The sample in **(a-d)** is 200-nm diameter fluorescent bead (FluoSpheres® Carboxylate-Modified Microspheres, 0.2 μm, dark red fluorescent (660/680), 2% solids). Scale bar in **(a)** is 300 nm.

Consequently, we can simulate out the Lorentzian shaped effective PSF of SAC fluorescence microscopy (for details, see **Supplementary Note 3** and **Supplementary Fig. 2**). Due to the similarities, the same theory used in STED [19] and GSD [13] microscopy is applicable to estimate effective resolution in SAC. When applying a greater intensity $I > I_S$, the optical resolution follows a square-root law given by:

$$\Delta r \approx \lambda / \left(2NA(1+P/P_s)^{1/2}\right) , \qquad (4)$$

where $P$ is the illumination power of the doughnut beam, given by $P = A*I$ ($A$ is the doughnut area,), $NA$ is the numerical aperture of the microscope objective and $\lambda$ is the excitation wavelength. At high saturations where $P \gg P_s$, the resolution $\Delta r$ decrease as $\sqrt{P/P_S}$.

***Experimental performance of SAC for various competition powers.*** Test experiments of 100-nm diameter fluorescent nanoparticles [T7279, TetraSpeck™ Microspheres, 0.1 μm, fluorescent 360/430 nm (blue), 505/515 nm (green), 560/580 nm (orange) and 660/680 nm (dark red)] are performed to evaluate the resolution of SAC microscopy, the results of which is shown in Fig. 3. Fig. 3(a,i) shows a standard fluorescence image of the nanoparticles using a confocal microscope, whereas Fig. 3(a,ii – a,vii) show the same set of particles being imaged with SAC microscopy using various input powers for the doughnut beam. The images clearly show that the acquired fluorescence intensity drops rapidly as the input beam power increases, which is also shown in Fig. 3(b), where the $P_S$ of this nanoparticle is indicated to be about 47 µW. Also, the expected sharpening of the central area with increasing powers of the doughnut is readily seen in Fig. 3(c). Notably, it can be seen that SAC sharply decreases the emitted fluorescence and $P_S$ is more than 20 times less than the counterpart of 10 mW typically required in CW STED. In addition, as is shown in Fig. 3(d), the greater the doughnut beam power $P$, the smaller the mean diameter of the bead. Measurements of the full-width at half-maximum (FWHM) values obtained from the images in Fig. 3(e,i) and drawn in Fig. 3(e,ii), indicate that our approach yields a 152 nm lateral resolution [Fig. 3(e)] at 240 µW, which is further verified by the frequency histogram of the fitted experimental bead size in Fig. 3(f). However, the experimental FWHM values are slightly larger than those predicted from Eq. (1). This discrepancy may have occurred due to unexpected dynamics of the absorption saturation, imperfections of the intensity null at the center of focused doughnut spot, and experimental misalignment. Discussions of similar appearance of deviations in GSD methods has been mentioned in the Supplemental Material of reference [20] as well.

The capacity to discern more details is further verified by the magnified views in Fig. 3(g). Two nanoparticles once fused into one spot in the confocal image are resolved in SAC images when we increase the competition power. It is clear that at an input beam power of 120 µW, the two particles are clearly resolved, and the distance between them is measured to be 175 nm as shown in Fig. 3(g,ii). Further increasing the doughnut beam power to 240 µW decreases the power in the region between the two particles, indicating that an increase in power improves the spatial resolution. Indicated in Fig. 3(h), the sub-diffraction resolution is also verified by a spatial-frequency distribution of SAC images containing additional high frequency components than that of confocal.

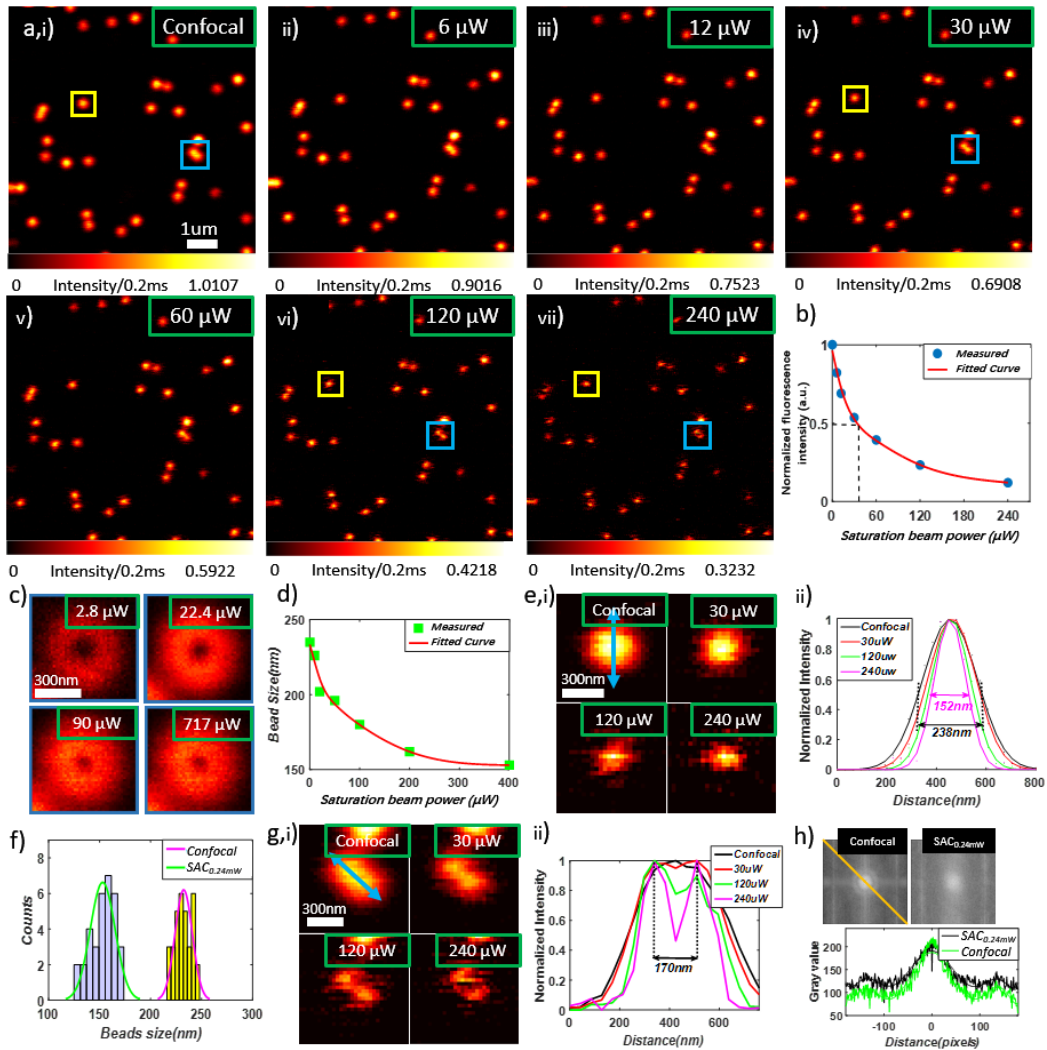

**FIG. 3. Comparison of experimental results obtained from conventional confocal microscopy and SAC microscopy. (a)** The 2D intensity distribution of 100-nm diameter fluorescent nanoparticles (field-of-view is 7.68 x 7.68 μm) obtained from **(i)** standard confocal microscopy using an input power of 5 μW, and **(ii-vii)** SAC microscopy as a function of input saturation-beam power (noted in the top right hand corner of each image). **(b)** The measured normalized fluorescence intensity as a function of the increase in doughnut beam power $P$. Inserted dashed lines reveal $P_s$, where the fluorescence drops down to half of its original amount. **(c)** Measured 2D intensity distribution of the saturation beam at various input powers. **(d)** Measured lateral FWHM of the detected fluorescence as a function of input saturation beam power; the trend appears to follow an inverse square law. Note that all the data are taken by counting the average value of Gaussian fitted beads sizes. **(e,i)** Magnified views of the fluorescence from a single bead in **(a,i,iv,vi and vii)** (indicated by yellow boxes). **(e,ii)** Intensity

profiles along the blue double arrow in **(e,i)**. **(f)** Histogram depicting the distribution of bead sizes in confocal **(a,i)** and SAC **(a,vii)** with the highest doughnut beam power. **(g,i)** Magnified views of two beads that appear clustered together in the confocal image **(a,i)** but can be distinguished in the SAC images **(a,iv, vi, vii)** (indicated by blue boxes). **(g,ii)** Intensity profiles along the blue double arrow line in **(g,i)**. **(h)** Top row shows corresponding 2D spatial frequency maps of confocal and SAC images in **(a,i)** and **(a,vii)**, while the bottom row shows the line profiles (along direction of yellow line shown in spatial frequency map) of these spectra. The pixel size is 30 nm with a dwell time of 0.2ms.

*Further applicability to imaging dense samples.* To evaluate the applicability of our approach to densely populated samples, two groups of 40-nm diameter fluorescent nanoparticles (F8789-FluoSphere Carboxylate-Modified Microspheres, 0.04 μm, dark red (660,680)) are imaged with SAC microscopy, the results of which are shown in Fig. 4. Considering the fact that the Lorentzian-shaped PSF of SAC contains significant low-frequency components (see in **Supplementary Fig. 2**), a deconvolution procedure will enhance the high frequency content in the image and hence the structural details [21]. The Richardson-Lucy algorithm is applied on the results shown in Figs. 4(a) and 4(b) with 10 iterations for a fair comparison, the results of which are shown in Figs. 4(c) and 4(d). The deconvolution method slightly improves the confocal resolution but merely narrows the size of spot instead of demonstrating structural details for densely packed areas. The SAC image, in contrast, is able to reproduce higher fidelity of these samples. We see from Fig. 4(c) that the confocal images can hardly discern the closely spaced nanoparticles due to the diffraction limit. As expected, SAC dramatically reduces the limit, the former blurred spots reveal eight distinct and homogenous spots in the final deconvolution result, which are further validated by the line profiles [Fig. 4(f)] along the white line in Fig. 4(e). Depicted in Fig. 4(h), line profiles across the sample images in Fig. 4(g) reveal a discernable distance of 100.5 nm which indicates that the resolving ability, represented by minimum discernable distance here, is ~100 nm, less than $\lambda/6$ ($\lambda$ =640 nm).

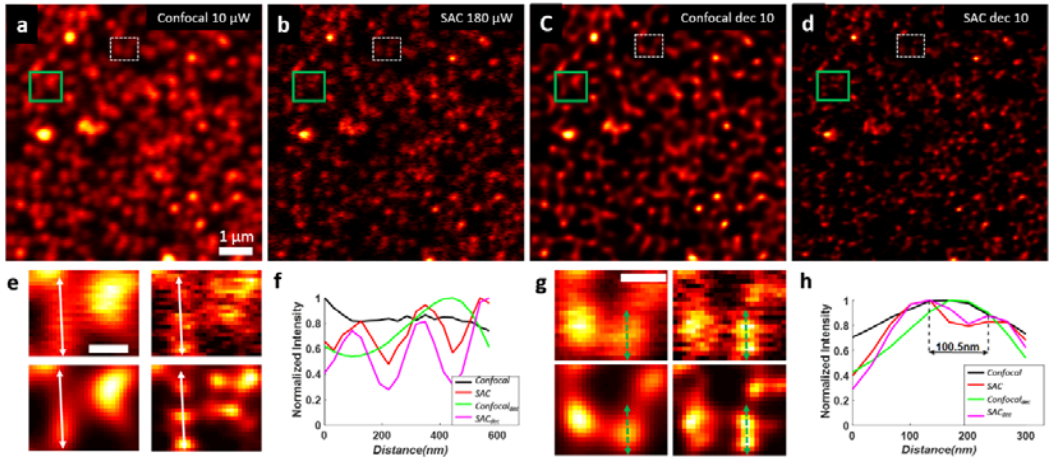

**FIG. 4. Imaging results of 40 nm fluorescent nanoparticles with confocal and SAC microscopy.** Images of **(a),** confocal at 10 μW, **(b),** SAC for saturation power of 180 μW, **(c),** Richardson-Lucy (RL) deconvolution result of **(a)** with 10 iterations, **(d),** RL deconvolution result of **(b)** with 10 iterations. **(e)** Magnified views of areas indicated by green boxes in **(a-d)**, scale bar, 300 nm. **(f)** Intensity profiles along the white lines in **(e)**. **(g)** Magnified views of areas indicated by write dashed boxes in **(a-d)**, scale bar, 300 nm. **(h)** Intensity profiles along the green lines in **(g)**. Pixel per size is 30 nm and the per dwell time is 0.5 ms.

In addition to imaging fluorescent nanoparticles, resolution enhancement in biological cells through SAC is also demonstrated in Fig. 5, where SAC was successfully applied to studying details of nuclear pore complex. Antibodies that appear clustered together in confocal images can be individually identified in SAC images.

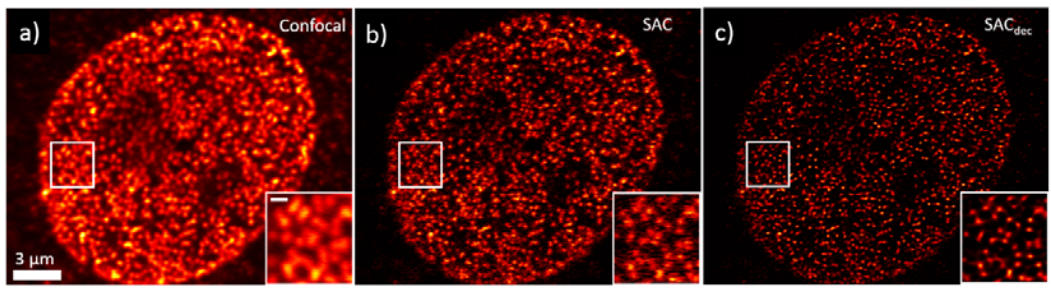

**FIG. 5. Imaging results of vero cells (Vero PFA, STAR-635P) with confocal and SAC microscopy.** Vero cell was stained with primary antibodies against the nuclear pore complex protein Nup153 and secondary antibodies conjugated with STAR-635P. Images of **(a),** Confocal at 1.5 μW, **(b),** SAC for saturation power of 40 μW, **(c),** RL deconvolution result of **(b)** with 20 iterations. Pixel per size is 30 nm and the per dwell time is 0.5 ms. Inset scale bar, 500 nm.

*Future prospects.* The super-resolution method proposed in this letter, can be deemed as a broadly applicable saturation and competition microcopy technique, which would not be restricted to fluorescence microcopy only. Obviously, the essence of SAC lies in the two key elements of generating a saturation effect to introduce competition and developing a method to extract the weak signal from the mixtures, which are of great inspiration to both fluorescent and non-fluorescent, yet saturable methods, for circumventing the diffraction limit.

*Summary.* In this letter, we developed a simple super-resolution microscopy method that manipulates the fluorescence absorption stage and demonstrated both synthetic and experimental results to showcase its applicability. Our results indicate that SAC can be realized with a single CW laser system, achieving a resolution lower than $\lambda/6$ with relatively low powers (about 200 µW) compared to similar super-resolution techniques such as CW-STED (requires ~60 mW) [18]. Also, compared to SAX, another lock-in based super-resolution technique, our proposed technique provides a better performance as it provides ~100 nm resolution compared to ~140 nm afforded by SAX [22]. In addition, experimental comparison with Structured Illumination Microscopy (SIM) upon biological cells [**Supplementary Fig. 5**] verifies its resolving ability. Though several factors could be considered to further improve the results shown in this work (see discussions in **Supplementary Note 4**), the already demonstrated gain in resolution is adequate for many applications, and thus opens new vistas for exploring simple methods to break the diffraction barrier via saturable mediums. Further, compared to the post-competition methods, e.g. STED, the equipment cost of SAC system is considerably cheaper. Considering the easy-setup, relatively low cost, illumination power and non-constraint of fluorophores used, we are confident that SAC can be further explored and widely applicable, thus facilitating optical observations.

## Acknowledgements

We thank the Abberior Instruments for the assistance of biological sample. This work was financially sponsored by the Natural Science Foundation of Zhejiang Province LR16F050001, the National Basic Research Program of China (973 Program) (2015CB352003), and the National Natural Science Foundation of China (61378051, 61427818, and 61335003).## References

[1]    E. Abbe, Archiv für mikroskopische Anatomie **9**, 413 (1873).

[2] S. W. Hell and J. Wichmann, Optics letters **19**, 780 (1994).
[3] W. A. Carrington, R. M. Lynch, E. Moore, G. Isenberg, K. E. Fogarty, and F. S. Fay, Science **268**, 1483 (1995).
[4] M. J. Rust, M. Bates, and X. Zhuang, Nature methods **3**, 793 (2006).
[5] E. Betzig, G. H. Patterson, R. Sougrat, O. W. Lindwasser, S. Olenych, J. S. Bonifacino, M. W. Davidson, J. Lippincott-Schwartz, and H. F. Hess, Science **313**, 1642 (2006).
[6] R. Heintzmann and C. G. Cremer, in *BiOS Europe'98* (International Society for Optics and Photonics, 1999), pp. 185.
[7] K. Fujita, M. Kobayashi, S. Kawano, M. Yamanaka, and S. Kawata, Physical Review Letters **99** (2007).
[8] C. B. Müller and J. Enderlein, Physical Review Letters **104**, 198101 (2010).
[9] J. Keller, A. Schönle, and S. W. Hell, Optics Express **15**, 3361 (2007).
[10] S. W. Hell, Nature Biotechnology **21**, 1347 (2003).
[11] D. Wildanger *et al.*, Advanced Materials **24**, 309 (2012).
[12] E. Rittweger, K. Y. Han, S. E. Irvine, C. Eggeling, and S. W. Hell, Nature Photonics **3**, 144 (2009).
[13] E. Rittweger, D. Wildanger, and S. W. Hell, Epl **86**, 14001 (2009).
[14] S. W. Hell and M. Kroug, Applied Physics B **60**, 495 (1995).
[15] P. Wang, M. N. Slipchenko, J. Mitchell, C. Yang, E. O. Potma, X. Xu, and J. X. Cheng, Nature Photonics **7**, 449 (2013).
[16] B. Yang, J. B. Trebbia, R. Baby, P. Tamarat, and B. Lounis, Nature Photonics **9**, 658 (2015).
[17] C. Eggeling, A. Volkmer, and C. A. Seidel, ChemPhysChem **6**, 791 (2005).
[18] K. I. Willig, B. Harke, R. Medda, and S. W. Hell, Nature Methods **4**, 915 (2007).
[19] V. Westphal and S. W. Hell, Physical Review Letters **94**, 143903 (2005).
[20] S. Bretschneider, C. Eggeling, and S. W. Hell, Physical Review Letters **98**, 132 (2007).
[21] R. Heintzmann, Micron **38**, 136 (2007).
[22] M. Yamanaka, Y. K. Tzeng, S. Kawano, N. I. Smith, S. Kawata, H. C. Chang, and K. Fujita, Biomedical Optics Express **2**, 1946 (2011).